# Non-ambiguity quantum teleportation protocol

Mario Mastriani

ORCID Id: 0000-0002-5627-3935

Teleportation is the most important and impactful tool in the arsenal of quantum communications with a particular projection on quantum internet. We propose a non-ambiguity alternative to the original teleportation protocol, which completely eliminates the classical-disambiguation-channel used by the original version. Experimental evidence on a quantum platform, via IBM cloud, is provided to demonstrate its performance.

*Introduction*.—Since the publication of the famous paper of Bennett et al [1], quantum teleportation has gained a central position in the world of quantum communications [2], for all that it represents and implies. This protocol consists of the following steps: (i) the generation and distribution of an entangled pair, in order to build a quantum channel between Alice and Bob. (ii) Alice receives the qubit to be teleported, which when interacting with one of the elements of the entangled pair gives rise to an ambiguous state, that is, a sum in which the original state is involved in four ways at once. (iii) Alice must make a measurement on that state, whose result is absolutely random. This measurement eliminates ambiguity, entanglement and also the original state, otherwise the No-Cloning Theorem [3] would be violated with each teleportation, which we know does not happen under any point of view. (iv) Alice transmits to Bob the result of the detection process via a classical channel, subject to the restrictions of the Einstein-Podolsky-Rosen (EPR) paradox [4], and therefore of the Special Relativity [5], which states that nothing can travel faster than light, implying that teleportation as a whole cannot be a process of instant transmission of useful information at all. (v) Bob applies a unit transformation based on the classical bits sent by Alice in order to reconstruct the teleported state.

The distribution of the entangled pair from which each teleportation process is inaugurated is also carried out through some physical means, usually using optical-fiber [6]. This procedure requires repeaters every approximately 50 km due to the optical properties of the fiber, for example: the absorption, the refractive index, and the reduced speed to which the entangled element travels (2/3 of speed of light in vacuum) which would make all teleportation efforts impossible over long distances by land without the use of quantum repeaters [7]. In some cases [8], an optical link based on a telescope-telescope pair is used, which does not require repeaters, as long as the eye contact between both elements is maintained. Teleportation uses two classical channels: the first one for the distribution of the entangled pair, and the second one between Alice and Bob to be used in the transmission of the measurement result made by Alice. Consequently, the elimination of this second channel, which represents the central idea of this work, does not imply in the least that the new teleportation protocol is instantaneous as a whole and thus collides with the Special Relativity [5], given that there is still the first classical channel for the distribution of the entangled pair, which gives rise to the quantum channel.

Based on what has been said so far, a question automatically arises: what is the impact of the second classical channel in the context of quantum communications? We will answer it with an example. If we use a quantum key distribution (QKD) protocol based on entanglement [9], we will have several channels at once: a first classical channel for the distribution of the entangled pair, which generates the quantum channel on which the public key will be teleported, the quantum channel generated as a result of the previous step, a second classical channel through which Alice transmits to Bob the result of her measurement, and a third classical channel on which the ciphered text travels, and which is encrypted by means of the teleported public key. If this architecture was intervened by a hacker, he would not be able to obtain the key that is teleported through the quantum channel, due to the fact that this channel is inaccessible to him, since an entanglement link is an intrinsically monogamous mean [10], that is, no third party can intervene without an initial consent of Alice and Bob, or what is the same, if at the time of the generation and distribution of the entangled elements no states of the Greenberger-Horne-Zeilinger (GHZ) or W [10] type were used, in other words, with full compliance from Alice and Bob, which, in this context, is absurd. The hacker can only intervene on the classical channels: the one that carries the entangled pair, the one that transmits the result of Alice's measurement, or the one that contains the encrypted message. If the hacker attacks the second classical channel, that is, that in which the result of Alice's measurement is sent to Bob, said hacker can alter the correct reconstruction of the teleported key and therefore the possibility to decrypt the message. In this case, the hacker cannot access the correct key but neither can Bob, taking into account that the key is fundamental to decipher the message information, i.e., to obtain the plain text. Hence the importance of the new protocol presented in this work. Finally, the impact that the new protocol will have on quantum Internet [11], both for quantum repeaters [7], and quantum swapping [12] is evident, giving rise to more robust (of greater immunity to noise), fast, and efficient procedures.

*Non-ambiguity quantum teleportation protocol.*—We will describe this protocol in a procedural way based on Fig. 1. First, an EPR pair of the kind $|\beta_{00}\rangle = 1/\sqrt{2}(|00\rangle + |11\rangle)$ is created and distributed between Alice and Bob from $t_2$. The horizontal green line separates Alice's and Bob's sides, and it can represent any arbitrary distance between them.

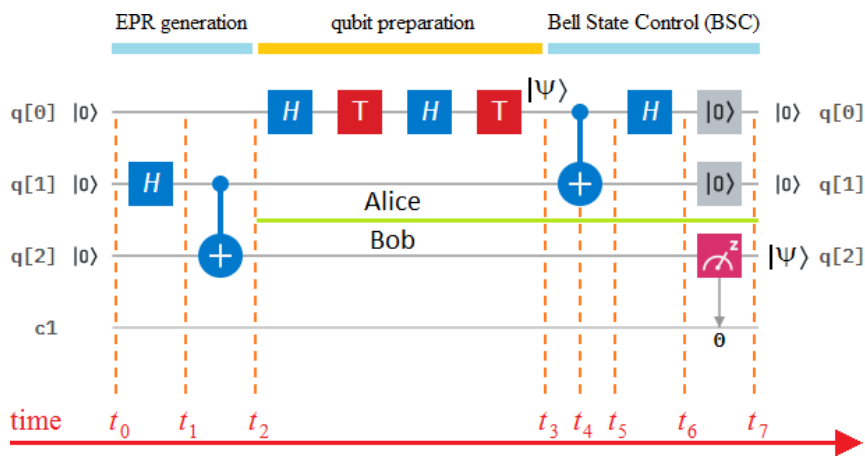

FIG. 1. Non-ambiguity quantum teleportation protocol using two qubit-reset gates [|0>] in gray.

A qubit to be teleported is prepared, for this case and without losing generality, from a combination of two gates, H and T: $|\psi\rangle = \begin{bmatrix} \alpha \\ \beta \end{bmatrix} = \text{HTHT}|0\rangle$, $T = \begin{bmatrix} 1 & 0 \\ 0 & \exp(i\pi/4) \end{bmatrix}$, $H = \frac{1}{\sqrt{2}}\begin{bmatrix} 1 & 1 \\ 1 & -1 \end{bmatrix}$, and $|0\rangle = \begin{bmatrix} 1 \\ 0 \end{bmatrix}$, $\exp(x) = e^x$, $e = 2.71828$, $\pi = 3.141592$, and $i = \sqrt{-1}$. The module on Alice's side, constituted by one CNOT, one H (Hadamard) and two qubit-reset gates [|0>], is called Bell State Control (BSC), and it is the one we will pay more attention to in the next development, where: $\text{CNOT} = \begin{bmatrix} 1 & 0 & 0 & 0 \\ 0 & 1 & 0 & 0 \\ 0 & 0 & 0 & 1 \\ 0 & 0 & 1 & 0 \end{bmatrix}$, and a qubit-reset gate [|0>] takes any arbitrary qubit to the ground state. As we will see next, it is evident that the qubit-reset gate [|0>] ∈ $\mathbb{C}^{2\times 2}$ is neither reversible nor unitary. For this protocol to work, we need that both elements of the EPR pair have the same polarity after applying this gate, specifically, the ground state |0>. Figure 2 shows an experiment which highlights the problem since once qubit-reset gate is applied to one of the two elements of an EPR pair, in this case q[0], both outcomes should simultaneously collapse to the ground state |0>, however, the problem is that each platform has its own version about how this gate should work, e.g., according to the simulators of IBM Q [13], Rigetti [14], and Quantum Inspire [15], this gate acts like a combination of: quantum measurement, *if-then-else* statement, and the inverter Pauli's *X* gate [16], as we can see in Fig. 3 for the IBM Q platform [13]. The problem with this implementation is that there is always a quantum measurement [17] before the

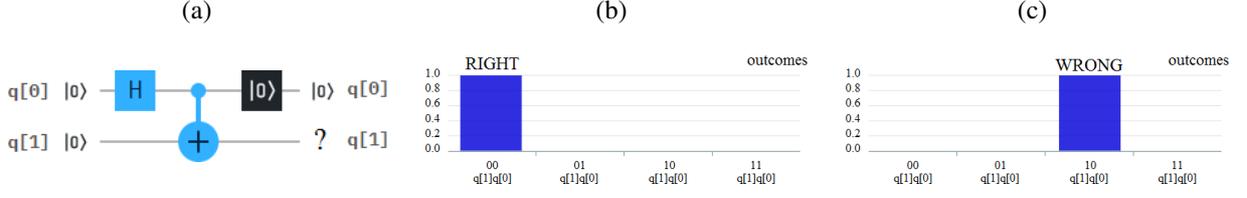

FIG. 2. Application of a qubit-reset gate on one of the elements of an EPR pair: (a) we need a ground state |0> in both outcomes simultaneously, however, q[1] will depends on the implementation type of this gate by the chosen platform, (b) the RIGHT outcomes, and (c) the WRONG outcomes.

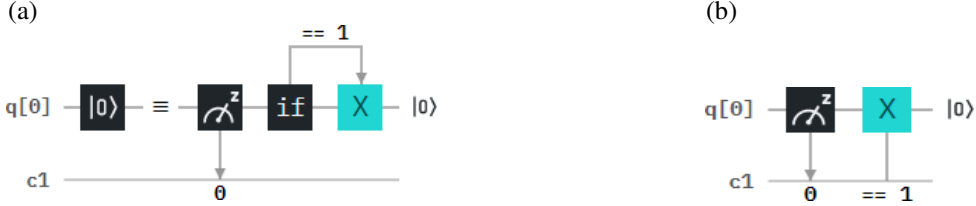

FIG. 3. Implementation of the qubit-reset gate for the IBM Q platform: (a) equivalence and schematic sketch, (b) typical implementation in the *drag-and-drop* circuit composer.

*if-then-else* statement with two possible and equally likely results: |0> and |1>, since if we want to apply this to a configuration like the one in Fig. 2(a), the measurement changes the result in both qubits at once making $|\beta_{00}\rangle$ collapse to |00> or |11>, and although q[0] finally ends in |0>, thanks the action of the *qubit-reset* gate, q[1] can be |1> half of the time. Therefore, this type of implementation is useless. To this problem another one is added, which is that it is currently not possible to implement the *if-then-else* statement on a Quantum Processing Unit (QPU)) [13, 14]. Then, the only options to implement *qubit-reset* gate efficiently are through:
a) a 90° polarizer [18-20], or vertical polarizer ($Pv$), and
b) a vertical polarizer $Pv$ with an amplification factor $A_f$ [18], i.e., $Av = A_f\, Pv = \sqrt{2}\, Pv$.
  There are exclusively two platforms that will allow us to implement these solutions:
- an optical quantum circuit [18], and
- Quirk simulator [21], which is the only one that allows designing gates even if they are not unitary.
  Therefore, for practical reasons, we will implement both options in Quirk [21] on one of the elements of an EPR pair. In Fig.4, the capital letters D, P, and B mean: density matrix, probability of |1> (where Off is equal to 0), and Bloch's sphere, respectively, while $Pv$ is a 90° polarizer, and $Av$ is a $Pv$ with an amplification factor $A_f$. $Pv$ eliminates all the *spin-down* outcomes, i.e., the |1>'s, only obtaining the 50% of the desired output, i.e., the |0>'s, while $Av$ always obtains the outcomes |00>. Therefore, the latter seems to be the best option. The matrices associated with these options are:

$$Pv = \begin{bmatrix} 1 & 0 \\ 0 & 0 \end{bmatrix} \qquad Av = A_f\, Pv = \sqrt{2}\begin{bmatrix} 1 & 0 \\ 0 & 0 \end{bmatrix} = \begin{bmatrix} \sqrt{2} & 0 \\ 0 & 0 \end{bmatrix}$$

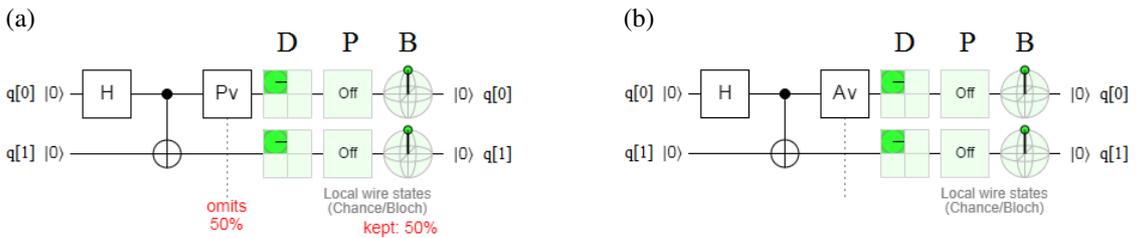

FIG. 4. Qubit-reset gate implementations in Quirk [21] via: a) a 90° polarizer ($Pv$), which eliminates all the *spin-down* out-comes, i.e., the |1>'s, only obtaining the 50% of the required output, i.e., the |0>'s. b) a $Pv$ with an amplification factor $A_f$ scheme ($Av$), which always obtains the outcomes |00>. The capital letters D, P, and B mean: density matrix, probability of |1> (where Off is equal to 0), and Bloch's sphere, respectively.

In Table I, we can see the time development of the protocol of Fig. 1 from $t_0$ to $t_7$, wherein a symbolic way "$\otimes$" is the Kronecker's product [16], which for simplicity [10], from now on, we will adopt $|x\rangle \otimes |y\rangle = |x\rangle|y\rangle$ in a generic form, while "$\underline{v}$" is the XOR operation, and $|+\rangle = H|0\rangle$.

We will only clarify in detail some sensitive instants of Table I, both on Alice's side and Bob's side, in order to better understand this protocol.

Table I. Protocol of Fig. 1 from $t_0$ to $t_7$.

| time ($t_i$) | qubits | | |
|---|---|---|---|
| | Alice's side | | Bob's side |
| | q[0] | q[1] | q[2] |
| 0 | $|0\rangle$ | $|0\rangle$ | $|0\rangle$ |
| 1 | $|0\rangle$ | $|+\rangle$ | $|0\rangle$ |
| 2 | $|0\rangle$ | $|\beta_{00}\rangle$ | $|\beta_{00}\rangle$ |
| 3 | $|\psi\rangle$ | $|\beta_{00}\rangle$ | $|\beta_{00}\rangle$ |
| 4 | $q[0]_4 = |\psi\rangle \otimes |\beta_{00}\rangle$ | $|\beta_{00}\rangle$ | $|\beta_{00}\rangle$ |
| 5 | $q[0]_5 = \text{CNOT}(|q[0]_4\rangle)$ | $|\beta_{00}\rangle$ | $|\beta_{00}\rangle$ |
| 6 | $q[0]_6 = H|q[0]_5\rangle$ | $|\beta_{00}\rangle$ | $|\beta_{00}\rangle$ |
| 7 | $|0\rangle$ | $|0\rangle$ | $|\psi_6\rangle \to X^0 Z^0 |\psi\rangle = |\psi\rangle$ |

*Alice's side:*

$t_3$: Alice gets an arbitrary and unknown state $|\psi\rangle$ to be teleported,

$$|\psi_3\rangle = |\psi\rangle = \alpha|0\rangle + \beta|1\rangle = [\alpha \quad \beta]^T \tag{1}$$

where $(\bullet)^T$ means *transpose of* $(\bullet)$, $|\alpha|^2 + |\beta|^2 = 1$, and $\alpha \wedge \beta \in \mathbb{C}$ of Hilbert's space [16], which constitutes a qubit on the Bloch's sphere with the following values: $\alpha = 0.9238$, and $\beta = 0.3826$, with probabilities: 0.854 for |0>, and 0.146 for |1>.

$t_4$: $|\psi\rangle = \alpha|0\rangle + \beta|1\rangle$ and $|\beta_{00}\rangle$ enter to a Bell State Control (BSC) module constituted by a *CNOT* gate, a *Hadamard's* (H) gate and two qubit-reset gates $[|0\rangle]$, however, the first interaction between both states only consists of a Kronecker's product. Then, a 3-partite state results in,

$$\begin{aligned}|\psi_4\rangle &= |\psi\rangle \otimes |\beta_{00}\rangle = |\psi\rangle|\beta_{00}\rangle = (\alpha|0\rangle + \beta|1\rangle)\tfrac{1}{\sqrt{2}}(|00\rangle + |11\rangle) \\ &= \tfrac{1}{\sqrt{2}}[\alpha|0\rangle(|00\rangle + |11\rangle) + \beta|1\rangle(|00\rangle + |11\rangle)] = \tfrac{1}{\sqrt{2}}[\alpha|000\rangle + \alpha|011\rangle + \beta|100\rangle + \beta|111\rangle] \\ &= [\tfrac{\alpha}{\sqrt{2}} \quad \tfrac{\beta}{\sqrt{2}} \quad 0 \quad 0 \quad 0 \quad 0 \quad \tfrac{\alpha}{\sqrt{2}} \quad \tfrac{\beta}{\sqrt{2}}]^T\end{aligned} \tag{2}$$

$t_5$: A *CNOT* gate is applied to $|\psi_4\rangle$,

$$|\psi_5\rangle = \frac{1}{\sqrt{2}}\big[\alpha|000\rangle + \alpha|011\rangle + \beta|110\rangle + \beta|101\rangle\big] \quad (3)$$

$$= \begin{bmatrix} \alpha/\sqrt{2} & 0 & 0 & \beta/\sqrt{2} & 0 & \beta/\sqrt{2} & \alpha/\sqrt{2} & 0 \end{bmatrix}^T$$

$t_6$: A *Hadamard's* (H) gate is applied to $|\psi_5\rangle$,

$$|\psi_6\rangle = \frac{1}{2}\big[|00\rangle X^0 Z^0|\psi\rangle + |01\rangle X^1 Z^0|\psi\rangle + |10\rangle X^0 Z^1|\psi\rangle + |11\rangle X^1 Z^1|\psi\rangle\big]$$

$$= \frac{1}{2}\big[|\beta_{00}\rangle X^0 Z^0|\psi\rangle + |\beta_{01}\rangle X^1 Z^0|\psi\rangle + |\beta_{10}\rangle X^0 Z^1|\psi\rangle + |\beta_{11}\rangle X^1 Z^1|\psi\rangle\big]$$

$$= \begin{bmatrix} \alpha/2 & \alpha/2 & \beta/2 & -\beta/2 & \beta/2 & -\beta/2 & \alpha/2 & \alpha/2 \end{bmatrix}^T \quad (4)$$

$$= \begin{bmatrix} \alpha/2 & 0 & \beta/2 & 0 & 0 & 0 & 0 & 0 \end{bmatrix}^T \to |\beta_{00}\rangle \to |00\rangle \to 00 \to X^0 Z^0$$

$$+ \begin{bmatrix} 0 & \alpha/2 & 0 & -\beta/2 & 0 & 0 & 0 & 0 \end{bmatrix}^T \to |\beta_{01}\rangle \to |10\rangle \to 10 \to X^0 Z^1$$

$$+ \begin{bmatrix} 0 & 0 & 0 & 0 & \beta/2 & 0 & \alpha/2 & 0 \end{bmatrix}^T \to |\beta_{10}\rangle \to |01\rangle \to 01 \to X^1 Z^0$$

$$+ \begin{bmatrix} 0 & 0 & 0 & 0 & 0 & -\beta/2 & 0 & \alpha/2 \end{bmatrix}^T \to |\beta_{11}\rangle \to |11\rangle \to 11 \to X^1 Z^1$$

where:

$$|\beta_{01}\rangle = \frac{1}{\sqrt{2}}\big(|0^A,0^B\rangle - |1^A,1^B\rangle\big),\ |\beta_{10}\rangle = \frac{1}{\sqrt{2}}\big(|0^A,1^B\rangle + |1^A,0^B\rangle\big),\ \text{and}\ |\beta_{11}\rangle = \frac{1}{\sqrt{2}}\big(|0^A,1^B\rangle - |1^A,0^B\rangle\big).$$

$t_7$: The two qubit-reset gates $[|0\rangle]$ of the BSC module of Fig. 1 eliminate the last three lines of Eq. (4), i.e., they eliminate the components on the bases $|\beta_{01}\rangle$, $|\beta_{10}\rangle$, and $|\beta_{11}\rangle$, or what is the same, those associated with spin-downs $|1\rangle$, allowing in this way to survive only the component with projection on the base $|\beta_{00}\rangle$ but on Bob's side, since on this side, i.e., Alice's side, we will only get spin-ups $|0\rangle$. That is, the action of both qubit-reset gates automatically eliminates the ambiguity represented by the last four lines of Eq. (4), the randomness associated with the measurement [17] of an entangled pair, entanglement, as well as, the original state $|\psi\rangle$ in order not to violate the No-Cloning Theorem [3].

*Bob's side:*

$t_7$: Therefore, the following and not ambiguous state results in,

$$|\beta_{00}\rangle \to X^0 Z^0 |\psi\rangle = |\psi\rangle. \quad (5)$$

As we can see, there is an abrupt jump from the state in which was the qubit q[2] of Fig. 1 in $t_6$ on Bob's side, i.e. $|\beta_{00}\rangle$, to the state obtained in Eq.(5) in moment $t_7$ by the action of the qubit-reset gates on the side of Alice, and without requiring in the least a classical channel of disambiguation to transmit the result of the measurement made by Alice. It is as if Alice always obtained a pair of zeroes 0 as a result of her measurement, therefore, if she always measures the same, she transmits the same and Bob receives the same thing, then, what transmit it for? Or else, why a disambiguation channel if there is no such disambiguation? The answer is evident, and it is the main reason by which the unitary transformation applied by Bob is reduced to the identity matrix.

*Experimental results*.—We prepare the qubit $|\psi\rangle$ to be teleported in q[0], which will be available from time $t_3$ of Fig. 1, with the following characteristics:

*Wavefunction* = (0.8535533906+0.3535533906j) |0> + (0.3535533906-0.1464466094j) |1>
*Amplitudes* = (0.8535533906+0.3535533906j) |0> + (0.3535533906-0.1464466094j) |1>
*Probability in* |0> = 0.8535533905932711, and
*Probability in* |1> = 0.14644660940672574

This previous analysis of the qubit to be teleported is essential to be able to compare it with the qubit teleported at the end of the protocol, i.e., qubit q[2] on the right of Fig. 1. Next, we are going to show the results of the new quantum teleportation protocol of Fig. 1 on Quirk [21] platforms.

First, we will begin with the implementation of Fig. 5, in which we use a pair of 90º polarizers *Pv*'s as qubit-reset gates. Both polarizers block the last three rows of Eq.(4), i.e., all terms of Eq.(4) with components |1>, transmitting only the 25% associated with the base $|\beta_{00}\rangle$, exclusivelly. The absolute coincidence between the two triads of metrics (D, P, B), i.e., those in q[0] after the qubit preparation procedure and those corresponding to q[2], at the end of the protocol, shows that teleportation has been perfect.

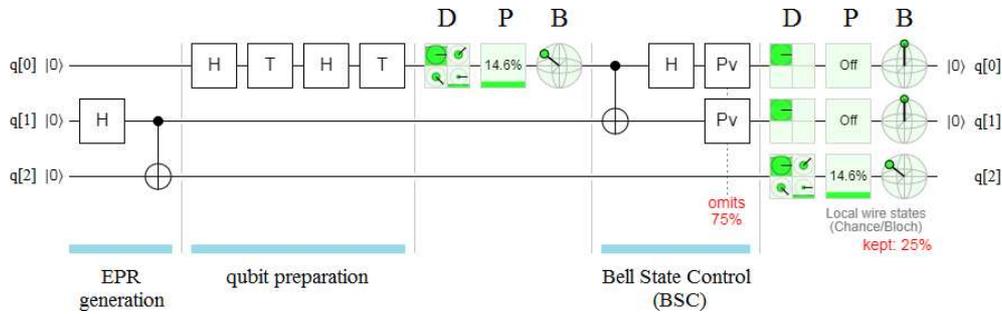

FIG. 5. Non-ambiguity quantum teleportation protocol in Quirk [21], using a 90º polarizer *Pv* as qubit-reset gate. Both polarizers block the last three rows of Eq.(4), i.e., all terms of Eq.(4) with components |1>, transmitting only the 25% associated with the base $|\beta_{00}\rangle$.

Figure 6 shows the other implementation of the new protocol where two *Pv* with amplification factors $A_f$ schemes *Av* are used as qubit-reset gates. Both *Av* convert all terms in Eq.(4) that have |1> to |0> without blocking any. It is evident that this last implementation is superior, being its results:

*Wavefunction* = (0.8535533906+0.3535533906j) |000> + (0.3535533906-0.1464466094j) |100>
*Amplitudes* = (0.8535533906+0.3535533906j) |000> + (0.3535533906-0.1464466094j) |100>
*Probability in* |000> = 0.8535533905932711, and
*Probability in* |100> = 0.14644660940672574

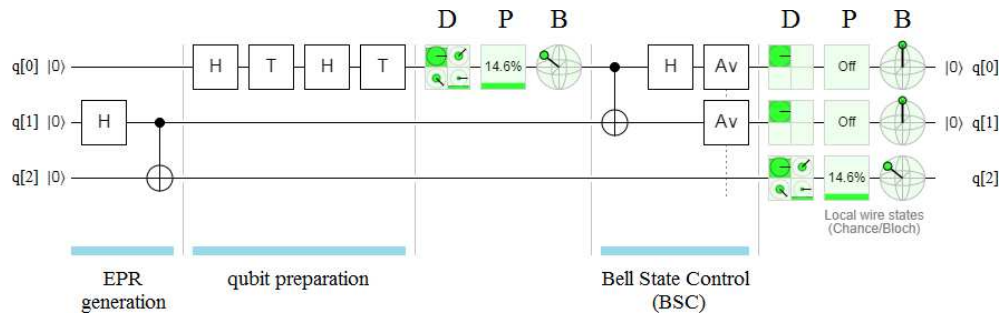

FIG. 6. Non-ambiguity quantum teleportation protocol in Quirk [21], using two *Pv* with amplification factors $A_f$ scheme *Av* as qubit-reset gate. Both *Av* convert all terms in Eq.(4) that have |1> to |0> without blocking any.

We can take these results to a more graphic representation like the one in Fig. 7, wherein (a) the height of the bar is the complex modulus of the wavefunction, (b) represents the probability bars or real part of the state for this experiment, (c) is the real part of the density matrix, and (d) is the imaginary part of the density matrix. As we can see, there is a total coincidence between the results of Fig. 7 and that of the qubit to be teleported, which evidences a perfect reconstruction of the teleported state. Moreover, the absolute cleanliness and the elegance of the results free us completely from performing complex arithmetic operations to finish reconstructing the state [22].

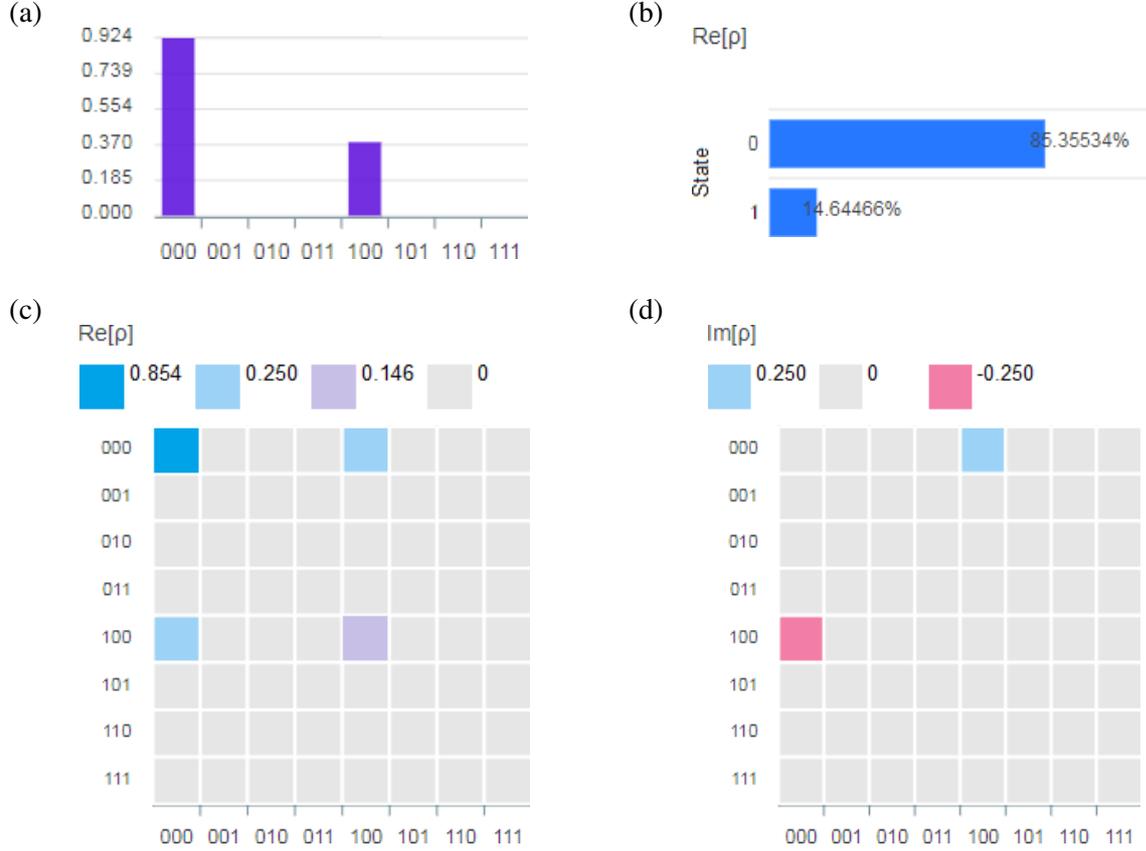

FIG. 7. Results of the teleportation of the state of Fig. 1. (a) The height of the bar is the complex modulus of the wavefunction. (b) The real part of the state. (c) The real part density matrix of the state. (d) The imaginary part density matrix of the state.

*Results of the new protocol teleporting a two-qubit state.*—In this case and in order to test the efficiency of the propose qubit-reset gate in dimensionally more complex teleportation protocols, we are going to teleport a two-qubit like,

$$|\psi\rangle = (0.854 + 0.354 j)|00\rangle + (0.354 - 0.146 j)|11\rangle, \tag{6}$$

thanks to a 3-qubit GHZ state like,

$$|GHZ_3\rangle = 1/\sqrt{2}(|000\rangle + |111\rangle). \tag{7}$$

The experiment is highlighted in Fig. 8, which begins with the generation and distribution of three entangled qubits thanks to a Hadamard (H) and two CNOT gates. We prepare a qubit like that of the previous experiment, however, the intervention of the CNOT gate at the output of the HTHT sequence generates a two-qubit like Eq.(6).

The 2-qubit Bell's State Control (BSC) module has three qubit-reset gates for this case, thanks to which Bob obtains two teleported qubits, in q[3] and q[4].

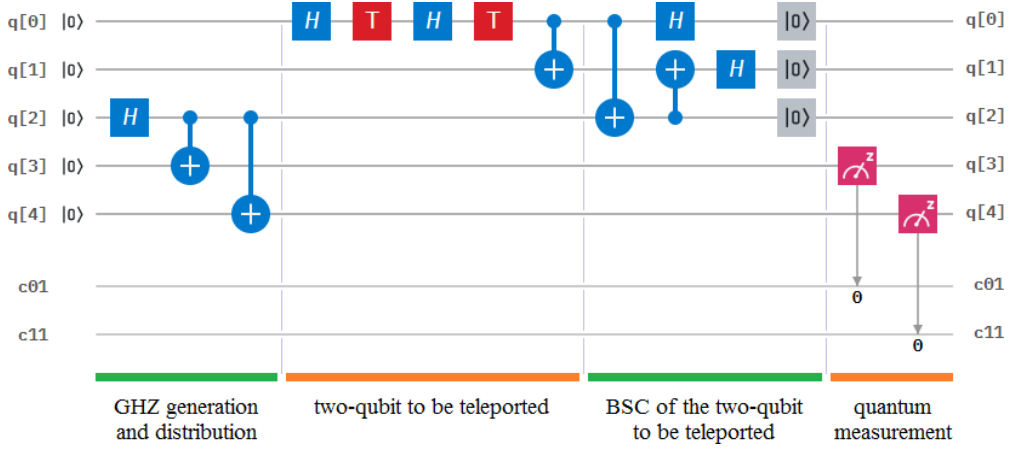

FIG. 8. Non-ambiguity quantum teleportation protocol using three qubit-reset gates [|0>] and a GHZ$_3$ state is used to teleport a two-qubit state.

Figure 9 shows the implementation of this experiment on Quirk [21], with the generation of the GHZ$_3$ entangled states between qubits q[2], q[3], and q[4]. The preparation of 2-qubit state to be teleported is constituted by the sequence of gates HTHT and the CNOT gate to its output, which affect the qubits q[0], and q[1]. After that, a 2-qubit BSC is employed with three $Pv$ with amplification factor $A_f$ ($Av$) to its output. The pair of triads (D, P, B) to the output of the qubit preparation module, qubits q[0] and q[1], perfectly coincides with those of the qubits q[3], and q[4], at the end of the protocol (right side) which evidences a perfect teleportation of the 2-qubit state.

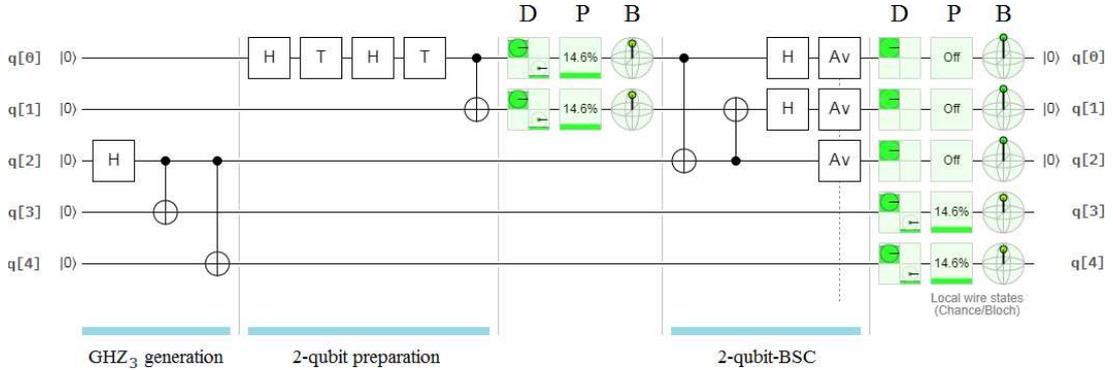

FIG. 9. Non-ambiguity quantum teleportation protocol in Quirk [21], using $Av$ scheme as qubit-reset gate. The three $Av$ convert the qubits q[0], q[1], and q[2] with |1> to |0> without blocking any.

The outcomes on the Quirk platform are: [ 0.854+0.354j "23 zeros" 0.354-0.146j "7 zeros" ], with (0.854+0.354j) for q[4]q[3]q[2]q[1]q[0]→00000, and (0.354-0.146j) for q[4]q[3]q[2]q[1]q[0]→11000, which constitute the expected successful outcomes for this experiment, in which, we have kept the same qubit to teleport of the previous experiment, to facilitate the comparison. Identical results can be obtained with another cloud access simulator called Quantum Programming Studio [23], thus demonstrating the performance of this protocol for dimensionally superior teleportations.

*Conclusion*.—We have successfully implemented a non-ambiguity quantum teleportation protocol: a teleportation protocol without a classical disambiguation channel, which makes instant teleportation, once the entangled pair has been distributed. Figures 6 and 7 show a complete coincidence between the state to be teleported and that teleported, which is a clear evidence of the proper functioning of the new protocol on Quirk [21], in fact, no need to perform any additional evaluation on the outcomes as is the case of the original protocol [22]. Similar results were achieved for the case of Fig. 9 in which a two-qubit, with similar characteristics to that of the experiment of Fig. 6, was successfully teleported.

These results apparently mean that it is possible an instant communication with the transmission of useful information based on entanglement, however, it is just an appearance, since the complete protocol implies the initial distribution of an entangled pair through a classical channel which is subject to the restrictions of Special Relativity [5]. Notwithstanding the above, the elimination of the classical disambiguation channel allows us: (i) a better Alice-Bob dialogue, which in a context of entanglement-based QKD represents a breakthrough to simultaneously protect the security and integrity of the data, and (ii) a much more robust protocol, i.e., of greater noise immunity. The new protocol interchangeably works with continuous variables as with qubits. Supplemental material contains the sources of Quirk [21] are available in [24].

Finally, it will be of fundamental importance to link the new protocol with both the purely European [25-28] as well as with the international [29, 30] effort of quantum internet, in order to evaluate the performance of this from a different perspective than the original.

*Competing interests*.—M.M. declares he has no competing interests.

*Funding*.—This work is funded by Qubit Reset Labs, an American start-up in quantum technology.

*Acknowledgements*.—M.M. would like to thank Antonio D. Corcoles-Gonzalez, and Jay Gambetta of IBM Q, as well as Tushar Mittal, Amy Brown, and Eric C. Peterson of Rigetti for their complete description of the possible implementation of the qubit-reset gate on their quantum physical machines in coherence time. Finally, a special acknowledgment to the board of directors of Qubit Reset Labs for its tremendous help and support.

*References*.—

1. C. H. Bennett, G. Brassard, C. Crépeau, R. Jozsa, A. Peres, and W. K. Wootters, *Teleporting an Unknown Quantum State via Dual Classical and Einstein-Podolsky-Rosen Channels*. Phys. Rev. Lett. 70, 1895 (1993)
2. P. E. Black, D. R. Kuhn, and C. J. Williams, *Quantum Computing and Communication: NIST* CreateSpace Independent Publishing Platform (2014)
3. W. K. Wootters, and W. H. Zurek, *A single quantum cannot be cloned*. Nature, 299, 802-803 (1982)
4. A. Einstein, B. Podolsky, and N. Rosen, *Can Quantum-Mechanical Description of Physical Reality Be Considered Complete?* Phys. Rev. 47:10, 777–780 (1935)
5. A. Einstein, H. A. Lorentz, H. Minkowski, and H. Weyl, *The Principle of Relativity: a collection of original memoirs on the special and general theory of relativity,* Courier Dover Publications. N.Y. (1952)
6. H. de Riedmatten, I. Marcikic, W. Tittel, H. Zbinden, D. Collins and N. Gisin, *Long distance quantum teleportation in a quantum relay configuration*, Phys. Rev. Lett., 92:4, 047904 (2004)
7. L. Jiang, J. M. Taylor, K. Nemoto, W. J. Munro, R. Van Meter, and M. D. Lukin, *Quantum repeater with encoding*, Phys. Rev. A 79, 032325 (2009)
8. T. Herbsta, T. Scheidl, M. Fink, J. Handsteiner, B. Wittmann, R. Ursin, and A. Zeilinger, *Teleportation of entanglement over 143 km*, PNAS, 112:46, 14202–14205 (2015)
9. A. K. Ekert, *Quantum cryptography based on Bell's theorem*. Phys Rev Lett, 67:6, 661–663 (1991)
10. R. Horodecki, P. Horodecki, M. Horodecki, and K. Horodecki, *Quantum Entanglement*, Rev. Mod. Phys. 81, 865 (2009)
11. M. Pant H. Krovi, D. Towsley, L. Tassiulas, L. Jiang, P. Basu, D. Englund, and S. Guha, *Routing entanglement in the quantum internet*, npj Quantum Information, 5:25 (2019)
12. T. Jennewein, G. Weihs, J.-W. Pan, and A. Zeilinger, *Experimental Nonlocality Proof of Quantum Teleportation and Entanglement Swapping*, Phys. Rev. Lett. 88, 017903 (2001)
13. https://quantum-computing.ibm.com/
14. https://www.rigetti.com/
15. https://www.quantum-inspire.com/
16. M. A. Nielsen, and I. L. Chuang, *Quantum Computation and Quantum Information*: 10[th] Anniversary Edition, Cambridge University Press, Cambridge (2011)
17. P. Busch, P. Lahti, J. P. Pellonpää, and K. Ylinen, *Quantum Measurement*. Springer, N.Y. (2016)
18. A. Furusawa, and P. van Loock, *Quantum Teleportation and Entanglement: A Hybrid Approach to Optical Quantum Information Processing*,Wyley-VCH, Weinheim, Germany (2011)
19. M. Sadiq, *Experiments with Entangled Photons: Bell Inequalities, Non-local Games and Bound Entanglement*, Ph.D. thesis, Department of Physics, Stockholm University, Malmö (2016)

20. C.K. Madsen, J.H. Zhao, *Optical fiber design and analysis: a signal processing approach*, John Wiley & Sons, N.Y. (1999)
21. https://algassert.com/quirk
22. M. Mastriani, *Is instantaneous quantum teleportation possible?* (2019) doi: 10.13140/RG.2.2.34116.99201/3, on-line available: https://www.researchgate.net/publication/337772422_Is_instantaneous_quantum_teleportation_possible
23. https://quantum-circuit.com/
24. M. Mastriani, Code in IBM Q and Quirk (2020) doi:10.13140/RG.2.2.15927.27049, on-line available: https://www.researchgate.net/publication/339075026_Suplementary_Material
25. W. Dür, R. Lamprecht, and S. Heusler, *Towards a quantum internet*, Eur. J. Phys., 38, 043001 (2017) 10.1088/1361-6404/aa6df7
26. H.J. Kimble, *The quantum internet*, Nature, 453, 1023–1030 (2008) 10.1038/nature07127
27. L. Gyongyosi and S. Imre, *Entanglement Access Control for the Quantum Internet*. arXiv:quant-ph/1905.00256v1 (2019)
28. L. Gyongyosi and S. Imre, *Opportunistic Entanglement Distribution for the Quantum Internet*. arXiv:quant-ph/1905.00258v1 (2019)
29. T. Satoh, S. Nagayama, T. Oka and R. Van Meter, *The network impact of hijacking a quantum repeater*, IOP Quantum Science and Technology, 3:3, 034008 (2018) doi:10.1088/2058-9565/aac11f
30. L. Gyongyosi, S. Imre, *Entanglement Accessibility Measures for the Quantum Internet*, Quantum Inf Process, 19:115 (2020) 10.1007/s11128-020-2605-y